\documentclass[twocolumn,notitlepage,superscriptaddress,showkeys,showpacs]{revtex4-1}
\usepackage{amsmath,amsfonts,amssymb,amsthm,epsfig,array}
\usepackage{dsfont}
\usepackage{graphics}
\usepackage{float}
\usepackage{verbatim}%\begin{comment} ... \end{comment}
\usepackage{color}
\usepackage[dvipsnames]{xcolor}
\usepackage[hidelinks,colorlinks,linkcolor=blue,
citecolor=blue,urlcolor=blue]{hyperref}
\usepackage[titletoc,title]{appendix}

\let\oldAA\AA
\renewcommand{\AA}{\text{\normalfont\oldAA}}

\newcommand{\mbt}{{MnBi$_2$Te$_4$~}}

\begin{document}
\title{Exchange Bias and Quantum Anomalous Hall Effect in the MnBi$_2$Te$_4$-CrI$_3$ Heterostructure}
\author{Huixia Fu}
\affiliation{Department of Condensed Matter Physics, Weizmann Institute of Science, Rehovot 7610001, Israel}
\author{Chao-Xing Liu}
\email{cxl56@psu.edu}
\affiliation{Department of Physics, the Pennsylvania State University, University Park, PA, 16802}
\author{Binghai Yan}
\email{binghai.yan@weizmann.ac.il}
\affiliation{Department of Condensed Matter Physics, Weizmann Institute of Science, Rehovot 7610001, Israel}
\begin{abstract}
The layered antiferromagnetic \mbt films have been proposed to be an intrinsic quantum anomalous Hall (QAH) insulator with a large gap. To realize this proposal, it is crucial to open a magnetic gap of surface states. However, recent experiments have observed gapless surface states \cite{Hao2019,Chen2019,Li2019arpes,Swatek2019}, indicating the absence of out-of-plane surface magnetism, and thus the quantized Hall resistance can only be achieved at the magnetic field above 6 T \cite{Deng2019,Liu2019,Ge2019}. In this work, we propose to induce out-of-plane surface magnetism of \mbt films via the magnetic proximity with magnetic insulator CrI$_3$. Our calculations have revealed a strong exchange bias $\sim 40$ meV, originating from the long Cr-$e_g$ orbital tails that hybridize strongly with Te $p$-orbitals. By stabilizing surface magnetism, the QAH effect can be realized in the MnBi$_2$Te$_4$/CrI$_3$ heterostructure. Our calculations also demonstrate the high Chern number QAH state can be achieved by controlling external electric gates. Thus, the MnBi$_2$Te$_4$/CrI$_3$ heterostructure provides a promising platform to realize the electrically tunable zero-field QAH effect.
\end{abstract}
\maketitle

%\tableofcontents

%\section{Introduction}
{\it Introduction: }
The quantum anomalous Hall (QAH) effect is a topological phenomenon characterized by quantized Hall resistance and zero longitudinal resistance~\cite{Haldane1988,Liu2016,chang2016quantum,wang2015quantum}. Different from the conventional quantum Hall effect, the QAH effect is induced by the interplay between spin-orbit coupling (SOC) and magnetic exchange coupling, and thus can occur in certain ferromagnetic (FM) materials at zero external magnetic field. Owing to its topological and dissipation-free properties, the QAH insulator is an outstanding quantum-coherent material platform for the next-generation quantum-based technologies, including spintronics ~\cite{Zutic2004,moore2010birth,vsmejkal2018topological} and topological quantum computations~\cite{Nayak2008}. Following the early theoretical predictions~\cite{Qi2006,Liu2008,Yu2010}, the QAH effect was first demonstrated in magnetically (Cr or V) doped  (Bi,Sb)$_2$Te$_3$ \cite{Chang2013,Checkelsky2014,Kou2014,Chang2015}, in which magnetic doping provides the required exchange coupling between magnetic moments and electron spins, and thus is essential for the occurrence of the QAH state. However, magnetic doping inevitably degrades sample quality with the presence of massive disorders and thus limits the critical temperature of the QAH state below 2K \cite{Chang2015}. Therefore, it is desirable to realize the QAH effect in intrinsic magnetic materials with stoichiometric crystals.

Recently, a tetradymite-type layered compound, MnBi$_2$Te$_4$, was proposed to be a promising topological material platform\cite{Gong2019,Li2019,Zhang2019,Otrokov2018MnBi2Te4} with intrinsic A-type anti-ferromagnetic (AFM) order, in which the magnetic moments of Mn atoms are ferromagnetically coupled within one septuple layer (SL) and anti-ferromagnetically coupled between the adjacent SLs, for the realization of the QAH effect, as well as other magnetic topological phases\cite{Qi2008TFT,Mong2010,Wan2011}. Early first principles calculations show that the QAH state can be realized in the MnBi$_2$Te$_4$ films with odd number of SLs at zero magnetic field for the ideal AFM order \cite{Li2019,Zhang2019,Otrokov2019}. The A-type AFM order was demonstrated via magnetization measurements for bulk MnBi$_2$Te$_4$ as the typical spin-flop transition was observed when the external magnetic field perpendicular to the SL plane was increased above 3.5T \cite{Otrokov2018MnBi2Te4,Lee2018,Deng2019,Yan2019,Wu2019,Chen2019Suppression}. However, the magneto-transport experiments in the MnBi$_2$Te$_4$ films only revealed a quantized Hall resistance for the magnetic field above 6T \cite{Deng2019,Liu2019,Ge2019}, larger than the critical field of spin-flop transition. Therefore, the thin film has already become ferromagnetic under this magnetic field. The predicted zero-field QAH state induced by the ideal AFM order has yet been demonstrated experimentally. The early angular-resolved photon emission spectroscopy (ARPES) measurements observed a band gap, ranging from 50 meV to hundreds of meVs \cite{Otrokov2018MnBi2Te4,Lee2018,vidal2019massive,chen2019searching}, of topological surface states (TSSs) in MnBi$_2$Te$_4$. However, this gap is shown to persist well above the N\'eel temperature and could be observed even at room temperature \cite{Lee2018,vidal2019massive,Otrokov2018MnBi2Te4}, making it unlikely originated from the AFM order. Indeed, more recent high-resolution ARPES studies based on synchrotron and laser light sources show that the TSS remains gapless below the N\'eel temperature \cite{Hao2019,Chen2019,Li2019arpes,Swatek2019}. 
The negligible magnetic gap of TSS is consistent with the absence of the zero-field QAH effect in magneto-transport measurements\cite{Lee2018,Deng2019,Liu2019,Wu2019, Ge2019}.
The absence of magnetic gap of TSSs suggests that the surface magnetism may not be well developed and different from the bulk AFM order. Physically, this is not surprising since more complex magnetic interactions, including dipole-dipole interaction and Dzyaloshinskii-Moriya interaction, may play an important role for the surface magnetic mechanism. Consequently, the surface Mn magnetic moments may be canted, or lie in the SL plane, or become disordered, all of which may lead to a gapless TSS. Furthermore, magnetic domains ubiquitously exist in AFM materials and cannot be easily eliminated even by field cooling. All these problems hamper the realization of zero-field QAH state in the MnBi$_2$Te$_4$ films. 

In this work, we propose to overcome the challenge of surface magnetism by coupling the \mbt films to a two-dimensional (2D) ferromagnetic insulator with the example of CrI$_3$ via exchange bias. Our density-functional theory (DFT) calculations on the \mbt/CrI$_3$ heterostructure show a ferromagnetic exchange bias around $40$ meV, much larger than the N\'eel temperature of \mbt (24 K \cite{Otrokov2018MnBi2Te4})  and the Curie temperature of CrI$_3$(61 K for bulk \cite{McGuire2015} and 45 K for monolayer\cite{Huang2017}) . Moreover, CrI$_3$ has little influence on electronic band structure of MnBi$_2$Te$_4$ films and thus the QAH state with the Chern number (CN) =1 can exist in 3- and 5-SL-thick \mbt, consistent with the early studies on pure \mbt films. We also studied the electric gating effect and the CrI$_3$/\mbt/CrI$_3$ heterostructures. Our results show (i) the high-CN QAH state with CN=3 can be achieved by tuning gate voltages and (ii) the strong exchange bias can always align the magnetization of both surfaces of \mbt films, thus driving even SL \mbt into the QAH state in the CrI$_3$/\mbt/CrI$_3$ heterostructure.

\begin{figure}[t]
\includegraphics[width=\linewidth]{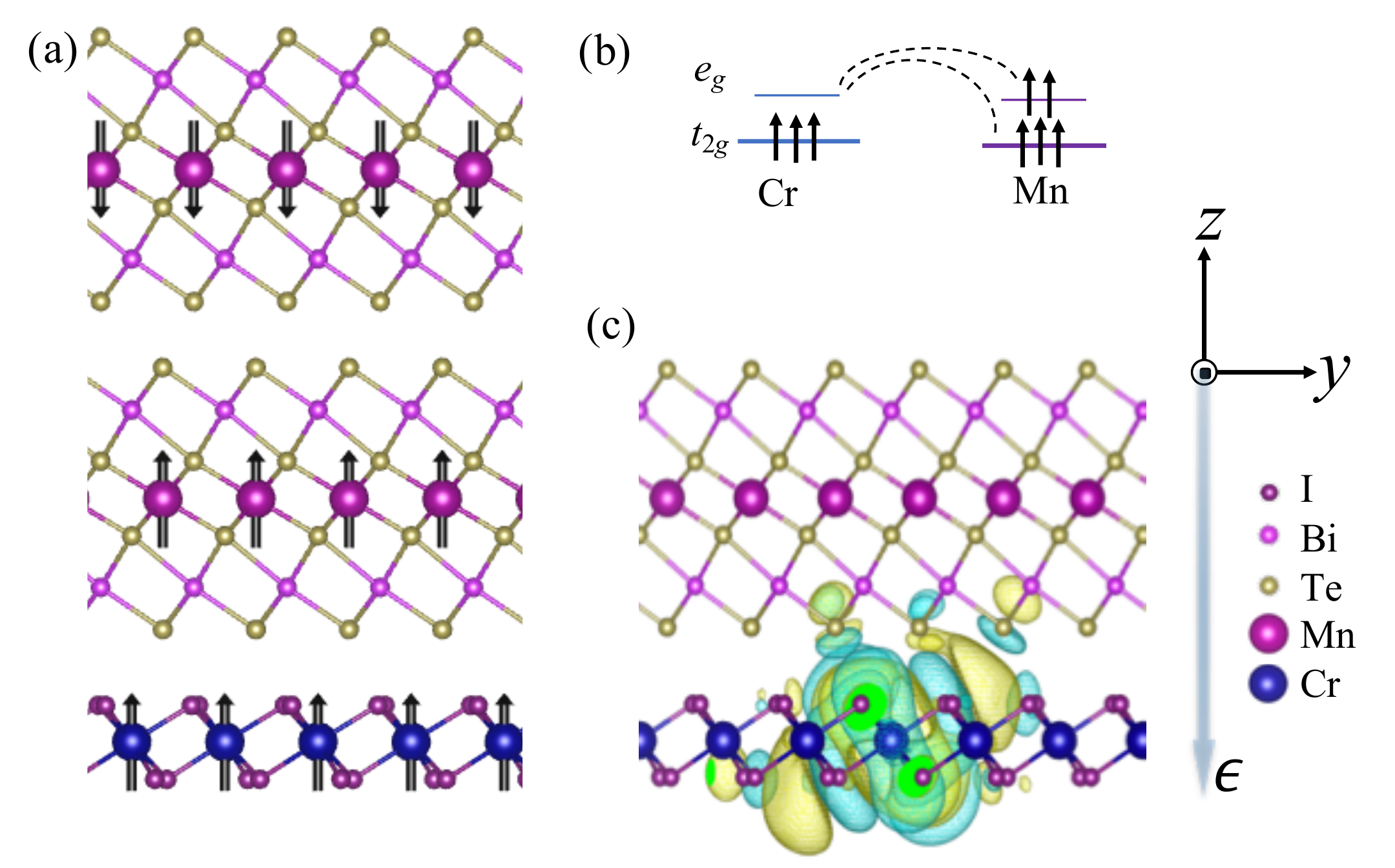}
\caption{\label{fig:structure}
The atomic structure of the MnBi$_2$Te$_4$/CrI$_3$ heterostructure.
(a) MnBi$_2$Te$_4$ and CrI$_3$ prefers the ferromagnetic type coupling while MnBi$_2$Te$_4$ layers remains the antiferromagnetic interaction. (b) Schematics of the exchange interaction between Cr-$e_g$ and Mn-$e_g t_{2g}$ states.
(c) The Wannier function of one of Cr-$e_g$ states. Its tails reach the neighboring Te atoms by crossing the van der Waals gap. Yellow and cyan colors represent positive and negative, respectively, isovalue surfaces of the Wannier function. An external electric field ($\epsilon$) along $-z$ can lift the energy of Cr-$e_g$ bands.
}
\end{figure}

{\it Ferromagnetic exchange bias at the MnBi$_2$Te$_4$/CrI$_3$ interface: } 
The required exchange bias material should provide strong magnetic coupling at the interface but not change the electronic states near the Fermi energy. Therefore, we choose a magnetic insulator, CrI$_3$\cite{Huang2017}. Its monolayer is ferromagnetic (FM) and can couple with  MnBi$_2$Te$_4$ through the van der Waals interface, which may disturb the band structure of MnBi$_2$Te$_4$ weakly. Because the interaction is determined by the interface layer, we only choose a monolayer of CrI$_3$ for the interface model. 

We construct interface models with one layer of CrI$_3$ and different layers of MnBi$_2$Te$_4$ on its top, as shown in Fig.~\ref{fig:structure}. Both materials share the same triangular lattice but different in-plane lattice parameters, $7.04 ~\AA$ for CrI$_3$ and $4.36 ~\AA$ for MnBi$_2$Te$_4$ from our DFT calculations, which is consistent with recent works\cite{Hou2019,Li2019,Zhang2018AIMnBi2Te4}. A $2\times 2$ supercell of CrI$_3$ can match well with a $3\times 3$ supercell of MnBi$_2$Te$_4$. Alternatively, the primitive unitcell of  CrI$_3$ can also match a $\sqrt{3}\times \sqrt{3}$ supercell of MnBi$_2$Te$_4$ with a mismatch of 7\%. Because we are mostly interested in the band structure of MnBi$_2$Te$_4$, we stretch the CrI$_3$ lattice to match the $\sqrt{3}\times \sqrt{3}$ MnBi$_2$Te$_4$ supercell. We fully optimized the atomic structures by including the van der Waals interactions in DFT calculations\cite{kresse1996} within the generalized-gradient approximation (GGA)\cite{perdew1996}. We have tested both models and found that they give similar results in the exchange coupling and band structure (see more information in the Supplementary Materials (SM) \cite{SMs}). Thus, we choose the smaller model, $\sqrt{3}\times \sqrt{3}$ MnBi$_2$Te$_4$/$1\times 1$ CrI$_3$, for further investigations in the following. 

At the interface, MnBi$_2$Te$_4$ exhibits strong the FM coupling with CrI$_3$. For 1 SL MnBi$_2$Te$_4$ on top of CrI$_3$, the energy difference between the FM and AFM coupling is about 40 meV. We note that different ways of stacking between two materials give very similar strength of exchange coupling, which is also true for the $3\times 3$ MnBi$_2$Te$_4$/$2\times 2$ CrI$_3$ case \cite{SMs}. When increasing the MnBi$_2$Te$_4$ layer to 2 SLs and more, the interface FM coupling remains with the same exchange energy and the two SLs couples still in the AFM way. Therefore, CrI$_3$ layer couples only with the neighboring MnBi$_2$Te$_4$ layer and does not affect the AFM order between different MnBi$_2$Te$_4$ layers. We point out that such an exchange coupling is much stronger than the magnitude of the exchange interactions between two MnBi$_2$Te$_4$ layers ($\sim$ 3 meV for $\sqrt{3}\times\sqrt{3}$ supercell) or two CrI$_3$ layers ($\sim$ 10 meV for $1\times 1$ unitcell\cite{Sivadas2018}). Therefore, CrI$_3$ can stably pin the FM order of the proximity MnBi$_2$Te$_4$ layer and act as an effective exchange bias. In addition, we find that SOC does not affect the magnetic coupling at the interface.

The strong exchange coupling originates in the orbital feature at the interface. The Mn site has $d^5$ configuration as $t_{2g}^3 e_g^2$ and the Cr site has $d^3$ as $t_{2g}^3 e_g^0$. There is a long exchange pathway from Cr-$e_g$ to Mn-$t_{2g}$ states through the intermediate I, Te, Bi and Te atoms, which is clearly beyond the simple super-exchange interaction. In the localized Wannier orbitals \cite{Mostofi2008}, we observe a crucial feature in the Cr-$e_g$ states. Tails of the Cr-$e_g$ Wannier functions extend beyond the van der Waals gap and strongly overlap with the neighboring Te-$p$ orbitals (see Fig.~\ref{fig:structure}c). This strong orbital overlap rationalizes the strong coupling between two materials. In addition, we note that AFM type coupling at the interface can also play a role of the exchange bias, although the present specific interface structure exhibits the FM coupling.

\begin{figure}[t]
\includegraphics[width=\linewidth]{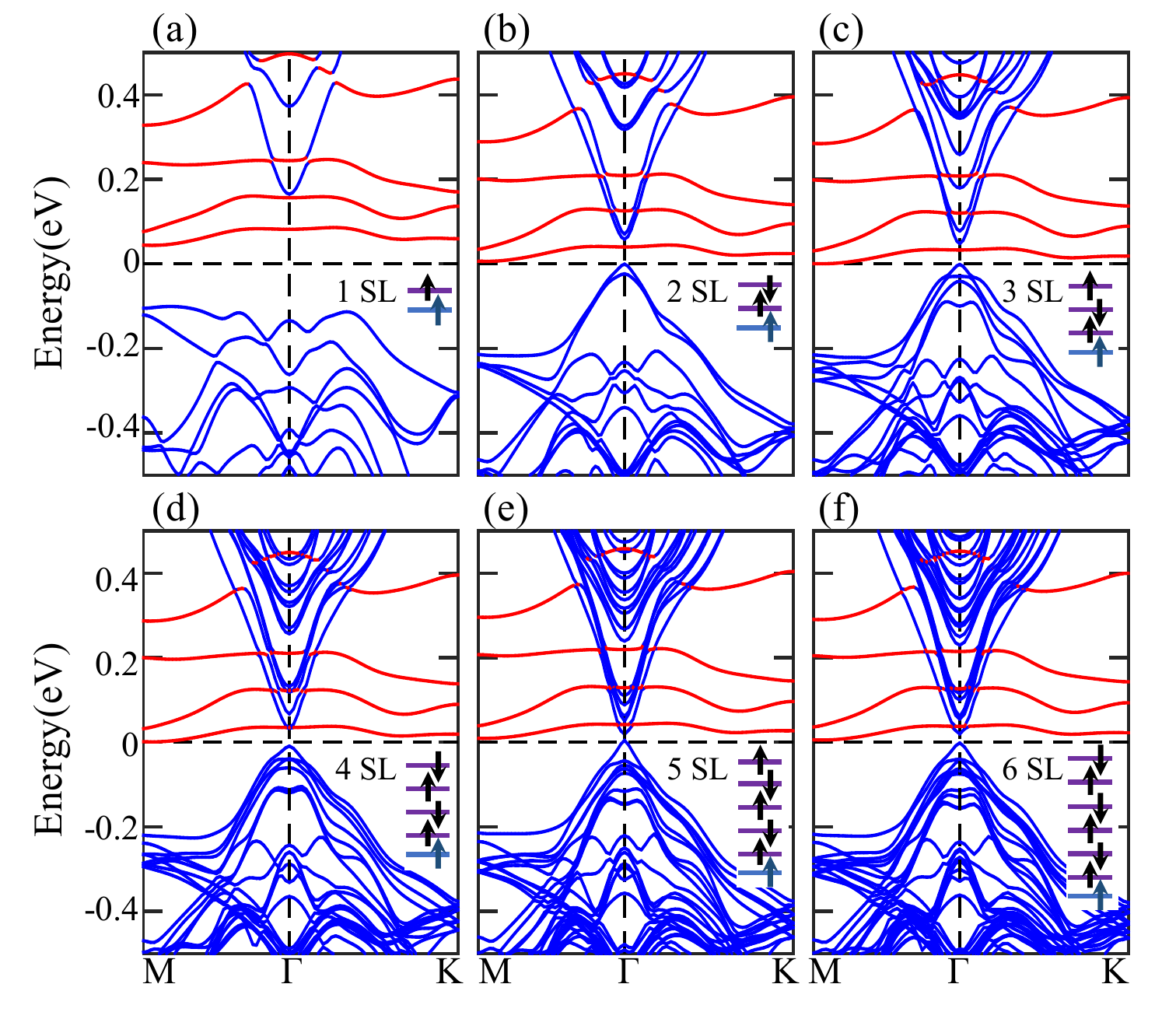}
\caption{\label{fig:band}
Band structures of \mbt~ thin films in proximity to a CrI$_3$ layer. From 1 to 6 - septuple layer (SL) - thick  \mbt~ are shown. Red lines indicate Cr-$e_g$ bands and blues ones for \mbt bands. The Fermi energy is set to zero.
}
\end{figure}

{\it QAH effect: }
We next investigate the electronic band structure and discuss its topological properties. Figure~\ref{fig:band} shows band structures for 1-6 MnBi$_2$Te$_4$ SL(s) on top of CrI$_3$. As discussed above, there is FM coupling between CrI$_3$ and neighboring the MnBi$_2$Te$_4$ SL and AFM coupling between MnBi$_2$Te$_4$ SLs. The interface band structure can be approximately regarded as an overlap of two different materials. An essential feature is the existence of an energy gap in these band structures, which is crucial for the realization of QAH effect. The occupied Cr-$t_{2g}$ bands are far below the valence bands of MnBi$_2$Te$_4$. The Cr-$e_{g}$ states overlap with the conduction band bottom of MnBi$_2$Te$_4$ and remain unoccupied. This means there is no charge transfer through the van der Waals junction. The calculated Cr-$t_{2g}$ and Cr-$e_{g}$ gap is about 1 eV, which is consistent with previous GGA calculations and can be corrected to about 1.5 eV by hybrid-functionals \cite{Zhang2015CrI3}. Although some Cr-$e_{g}$ bands appear as the lowest conduction bands at the interface for thinner \mbt films (1-4 SLs), they do not affect our understanding of the band structure topology. When the \mbt layer is thicker (e.g. 5-6 SLs), the \mbt states become the lowest conduction bands. Thus, CrI$_3$ serves an ideal proximity exchange bias without destroying the \mbt band structure.

We find that isolate MnBi$_2$Te$_4$ layers are trivial magnetic insulators for 1, 2, 4 and 6 SLs thick and QAHE insulators for 3 and 5 SLs, which is consistent with recent theoretical studies~\cite{Li2019,Otrokov2019}. Here, the QAH insulator possesses the CN = 1, as showed by our Berry phase calculations using the Wilson loop method~\cite{Yu2011,Soluyanov2011} in the 2D Brillouin zone. In proximity to the CrI$_3$ layer, \mbt band structures are modified weakly without changing their topological nature. For example, the isolated \mbt layer of 2, 4 or 6 SLs thick exhibits the double degeneracy in the band structure caused by the symmetry combining spatial inversion and time reversal. The existence of the CrI$_3$ layer breaks weakly this symmetry and splits the degenerate bands. We verify the topological character of the interface structures by observing the band gap evolution with respect to the SOC strength. For 3 and 5 SLs thick \mbt/CrI$_3$, the band gap closes at about 90\% of the normal SOC strength but re-opens an energy gap as increasing SOC, showing a topological phase transition (TPT)\cite{SMs}. The QAH insulator gaps are 49 and 14 meV for the 3 and 5 SLs interface, respectively. For 1, 2, 4 and 6 SLs thick \mbt / CrI$_3$, however, the band gap remains open as varying SOC from 0 to 100\%. (see more information in Ref.~\cite{SMs}).

\begin{figure}[t]
\includegraphics[width=\linewidth]{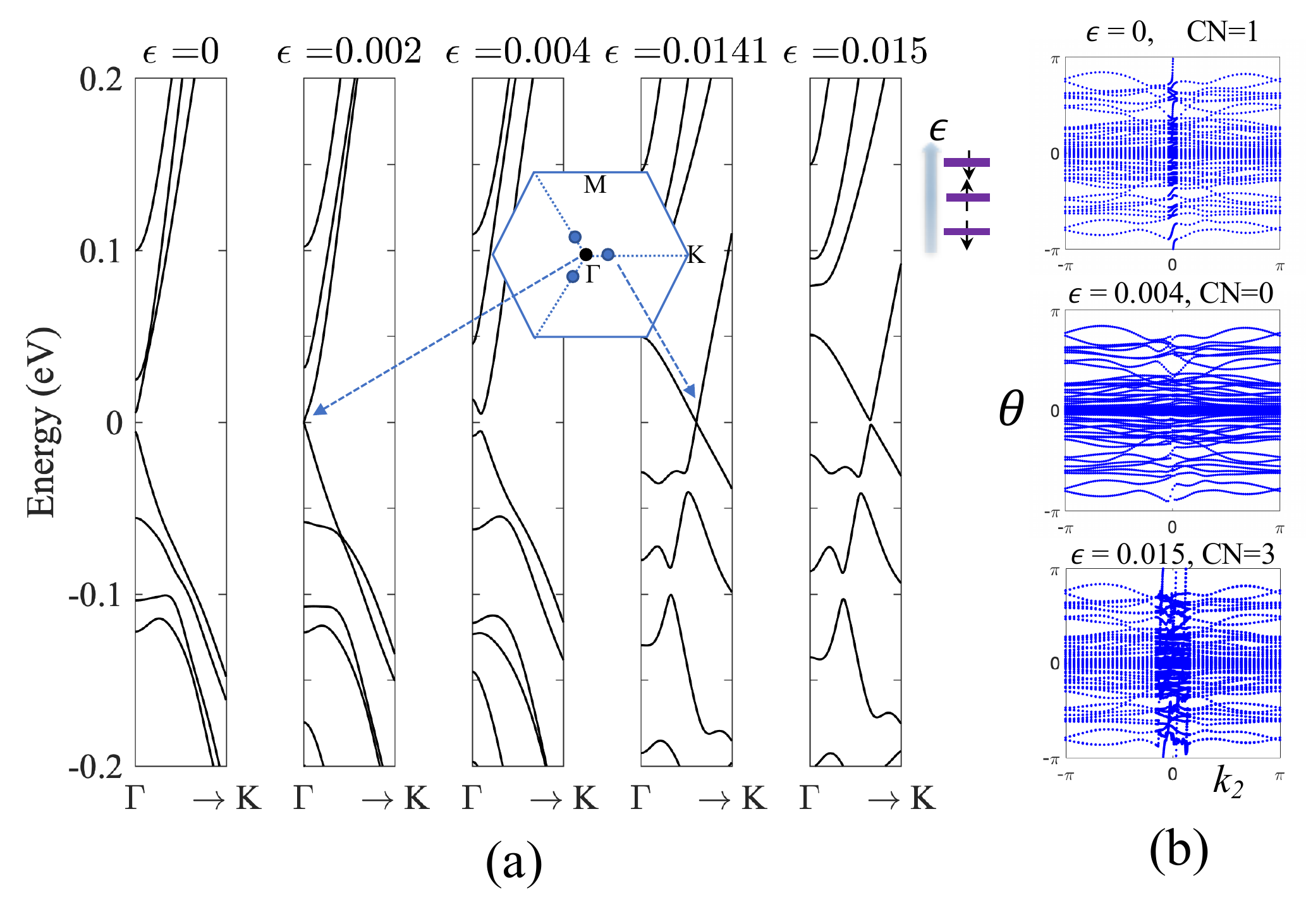}
\caption{\label{fig:efield}
(a) Band structure evolution of a 3-SL-thick as increasing the electric field ($\epsilon$). Two topological transitions occur at $\epsilon=0.002$ at the $\Gamma$ point and 0.0141 V/\AA ~along $\Gamma - K$ lines.
(b) The Berry phase $\theta$, i.e. the Wannier charge center, accumulated along a Wilson loop ($k_1 \in [-\pi, \pi]$) as varying $k_2$. It is topologically equivalent to the edge state spectra. Corresponding Chern numbers (CN) are one, zero and three for $\epsilon=0.00, 0.004$ and 0.015 V/\AA, respectively. 
}
\end{figure}

{\it Electrically tunable high-Chern-number QAH effect: }
The 2D layered structure offers an opportunity to tune the band structure topology by applying a vertical electric field. The electric field induces different potential variation at different layers and subsequently modifies the overall band structure and its topological nature. 
For the interface structure, an electric field ($\epsilon$) along the $-z$ direction can push the Cr-$e_g$ states up in the conduction band, as illustrated in Fig.~\ref{fig:structure}b, leaving only \mbt states right above and below the energy gap. Further increasing the electric field can induce an inversion between the occupied and unoccupied bands, giving rise to the TPT. Since the CrI$_3$ brings little modifications to the low-energy band structure of \mbt, we only consider isolated \mbt models when applying an electric field in following discussions.

The electric field can induce the high-CN QAH state. In a simple two-band model\cite{Qi2006}, a band inversion at the $\Gamma$ point usually leads to a change of the CN by $\pm 1$. If the band inversion occurs at generic k-points, it can induce a jump of the CN by the number of the transition points. The \mbt film under an electric field exhibits two important symmetries, the three-fold rotation (denoted as $C_3$) and a combined symmetry between the time-reversal and mirror reflection (denoted as $TM$). Since the mirror plane crosses the $\Gamma-M$ line in the Brillouin zone and perpendicular to the layer plane, the $\Gamma-K$ line is invariant under the $TM$ symmetry. Therefore, if a transition happens at a generic k-point away from the $\Gamma-K$ line, the gapless points must exist at six different k-points related by the $C_3$ and $TM$ symmetries. If a transition happens along the $\Gamma-K$ line that is invariant under $TM$, the gapless points must simultaneously appear at three different k-points related by $C_3$ (See the inset of Fig. \ref{fig:efield}a). If a transition appears at $\Gamma$ that is invariant under both $C_3$ and $TM$ symmetries, a single Dirac point transition can occur. 

To verify this scenario, we carried out band structure calculations on 3-SL-thick \mbt and demonstrate that the CN can jump by both 1 and 3 via applying a small electric field, as shown in Fig.~\ref{fig:efield}. At zero electric field, the 3-SL-thick \mbt is a QAH state with CN 1 and changes to a trivial insulator for $\epsilon=0.005~V/\AA$. This transition is through a gap-closing point at $\Gamma$ for $\epsilon=0.002/\AA$. For a larger electric field ($\epsilon=0.015$ V/\AA), another transition occurs with three gap-closing points along the $\Gamma-K$ lines, leading to a QAH state with CN = 3.
Furthermore, the electric field can also drive the \mbt film with even number of SLs from a trivial magnetic insulator with zero CN to the QAH state. For instance, $\epsilon=0.03~V/\AA$ induces a TPT with three gapless points along the $\Gamma-K$ lines in the 2-SL-thick \mbt film at $\epsilon=0.023~V/\AA$, resulting in the QAH state with CN = 3 (See more information in SM \cite{SMs}).

\begin{figure}[t]
\includegraphics[width=\linewidth]{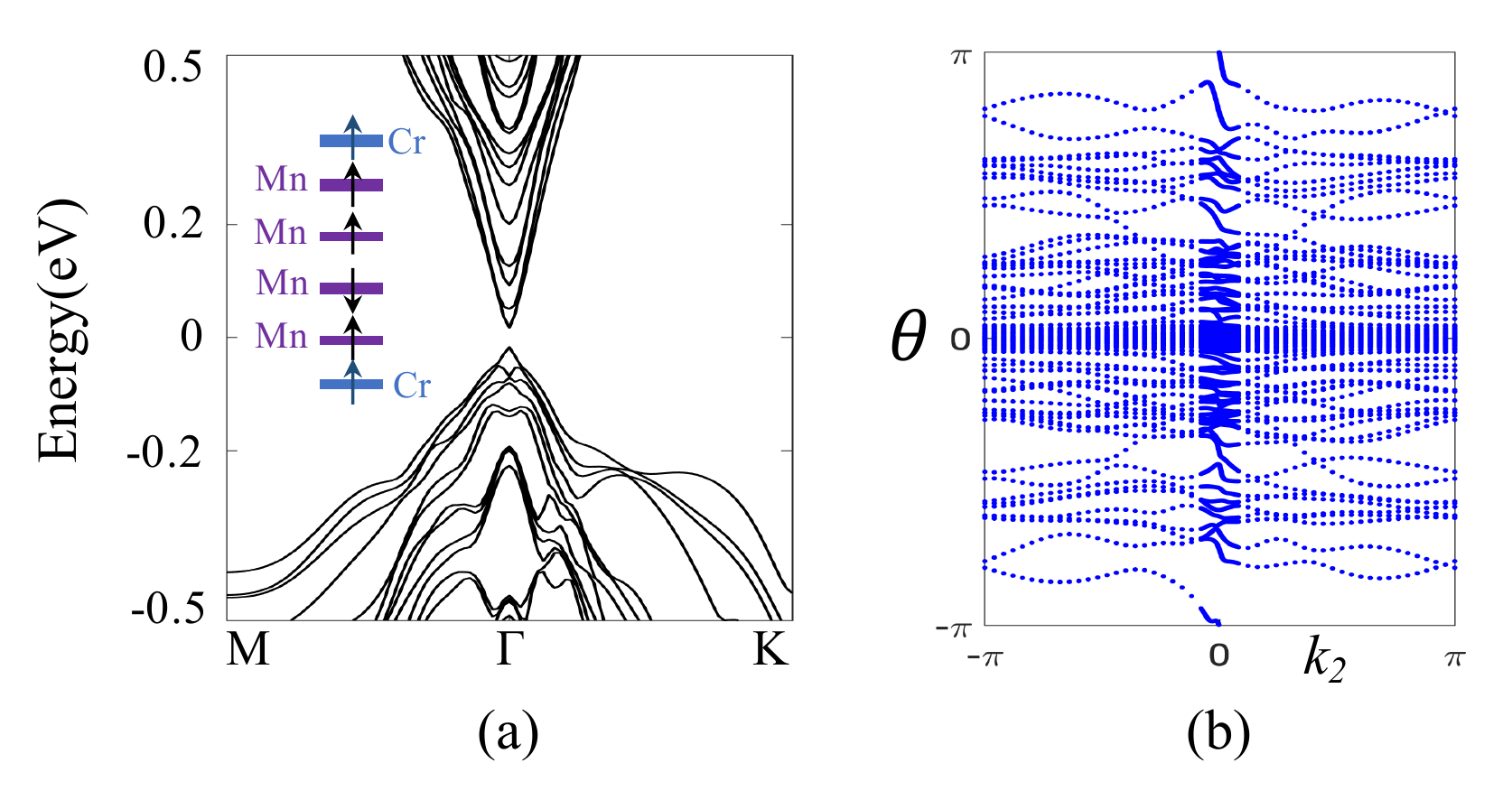}
\caption{\label{fig:sandwich}
(a) Band structure of a 4-SL-thick \mbt~ with a special magnetic order. When it was sandwiched between two CrI$_3$ layers, the antiferromagnetic type coupling between \mbt~ layers are changed by the top CrI$_3$ layer, as illustrated in the inset. (b) Corresponding Berry phase of the Wilson loop, with a Chern number CN = --1. 
}
\end{figure}

{\it Sandwiched \mbt structures: } Given the short range nature of exchange bias, the CrI$_3$ is expected to align the magnetization of the bottom \mbt layer in the \mbt/CrI$_3$ heterostructure, but may have little influence on the top \mbt layer when the film thickness is large. This issue can be resolved by considering a sandwiched structure CrI$_3$/MnBi$_2$Te$_4$/CrI$_3$. For the \mbt films with an odd number of SLs, the AFM order in \mbt is compatible with the FM orders in the top and bottom CrI$_3$ monolayers. In contrast, for the \mbt films with an even number of SLs, the compensated AFM ordering between \mbt layers can be changed by CrI$_3$. As an example of the 4-SL case in Fig.~\ref{fig:sandwich}, the magnetization of the top \mbt SL feels frustration from the upper CrI$_3$ layer and the lower \mbt layer. Because the \mbt-CrI$_3$ coupling is much stronger than the \mbt-\mbt coupling, magnetic moments of the top \mbt SL aligns parallel to those of the CrI$_3$ layer. Such a re-arrangement of magnetic moments in the \mbt SL leads to a net magnetization for the \mbt film. As verified by our band structure calculations, reversing magnetic moments of the top \mbt SL layer is indeed energetically favored by $\sim$ 40 meV. Subsequently the system becomes a QAH state with an energy gap of 34 meV. As shown by the Wilson loop calculations, it exhibits a nontrivial CN $=-1$. Therefore, the sandwich configuration may always provide a QAH insulator for either odd or even number of \mbt SLs.

{\it Discussion and Conclusion: } In summary, the magnetic order of MnBi$_2$Te$_4$ thin film can be pinned and also manipulated by a strong exchange bias in proximity to CrI$_3$. Thus, the heterostructures with \mbt and CrI$_3$ provide an experimentally feasible platform to realize the QAH effect. An external electric field can further modify the thin film band structure and induce QAH effect with large CNs. Since the magnetic insulator CrI$_3$ disturbs weakly the electronic states of \mbt, it can also be used to pin the surface magnetic order of the bulk \mbt and assist the observation of the axion insulator phase\cite{Qi2008TFT,Mong2010} in ARPES. Additionally, it is worth noting that other magnetic insulators with out-of-plane magnetization, such as Tm$_3$Fe$_5$O$_{12}$ (TmIG)\cite{Avci2017} and Cr$_2$Ge$_2$Te$_6$\cite{Gong2017}, may also play the same role of exchange bias as CrI$_3$.

{\it Acknowledgement: }
We acknowledge helpful discussions with Cui-zu Chang and Xiaodong Xu. 
Work at Penn State (CXL) was primarily supported by the U.S. Department of Energy (DOE), Office of Science, Basic Energy Sciences (BES) under Award DE-SC0019064. CXL also acknowledges the support from the Office of Naval Research (Grant No. N00014-18-1-2793) and Kaufman New Initiative research grant KA2018-98553 of the Pittsburgh Foundation. B.Y. acknowledges the financial support by the Willner Family Leadership Institute for the Weizmann Institute of Science, the Benoziyo Endowment Fund for the Advancement of Science,  Ruth and Herman Albert Scholars Program for New Scientists, the European Research Council (ERC) under the European Union's Horizon 2020 research and innovation programme (Grant No. 815869).

%\bibliography{bibfile_references}

\clearpage
\appendix

% Add 'S' to the numbering inside the supplement
\renewcommand{\thesection}{\arabic{section}}
\renewcommand{\thetable}{S\arabic{table}}
\renewcommand{\thefigure}{S\arabic{figure}}
\renewcommand{\theequation}{S\arabic{equation}}
\setcounter{section}{0}
\setcounter{figure}{0}
\setcounter{table}{0}
\setcounter{equation}{0}

\section{Calculation Methods}
Density functional theory (DFT) calculations were performed using the Vienna ab initio simulation package (VASP) with core electrons represented by the projector-augmented-wave (PAW) potential.  Plane waves with a kinetic energy cutoff of 270 eV were used as the basis set.  Geometry optimization was carried out until the residual force on each atom was less than 0.01 eV/\AA. 
We projected the Wannier functions of the bulk MnBi$_2$Te$_4$ in the AFM phase. Based on the bulk tight-binding parameters of Wannier functions, we constructed the slab model for MnBi$_2$Te$_4$ thin films and evaluated their band structures and Berry phases.

\begin{figure}[hb]
\includegraphics[width=\linewidth]{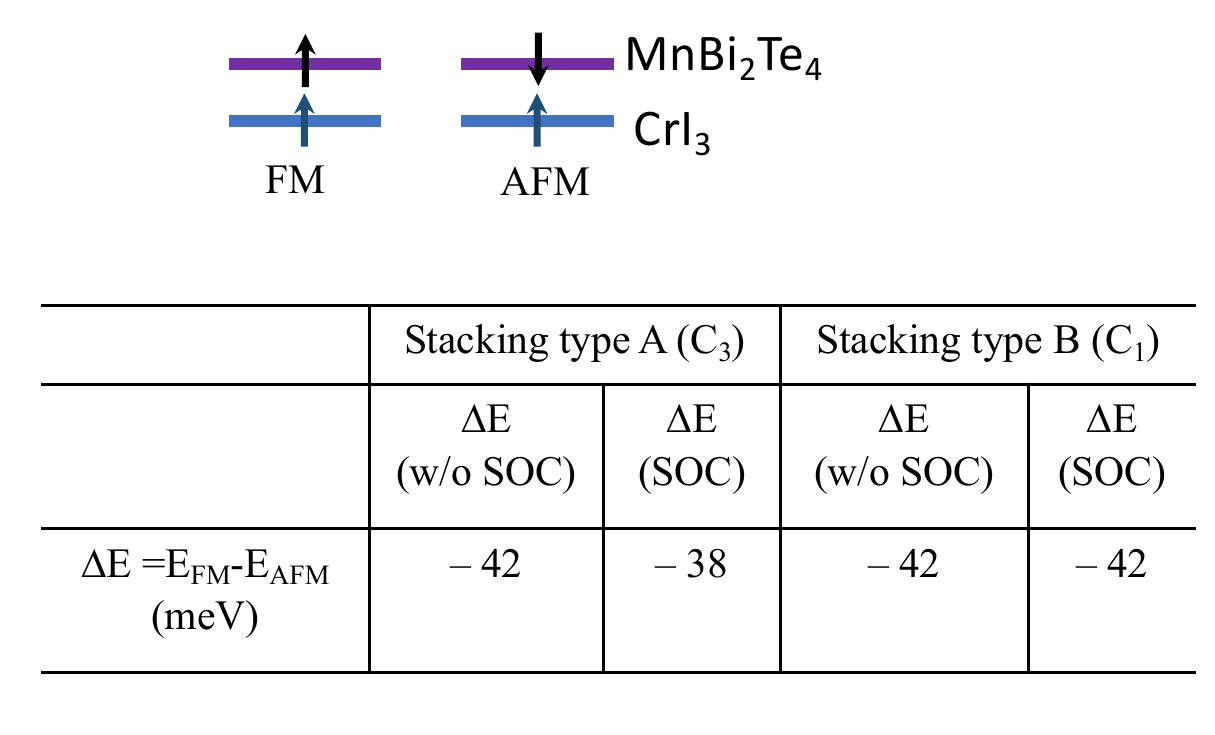}
\caption{\label{fig:1SLstacking}
The magnetic coupling energy $\Delta E$ for different stacking way of the interface for 1-SL \mbt on CrI$_3$. The type-A stacking has $C_3$ rotational symmetry while the type-B has no. Results with and without spin-orbit coupling (SOC)  are shown. Different way of stacking and SOC as well do not sensitively modify the exchange coupling at the interface. 
 }
\end{figure}
\begin{figure}[hb]
\includegraphics[width=\linewidth]{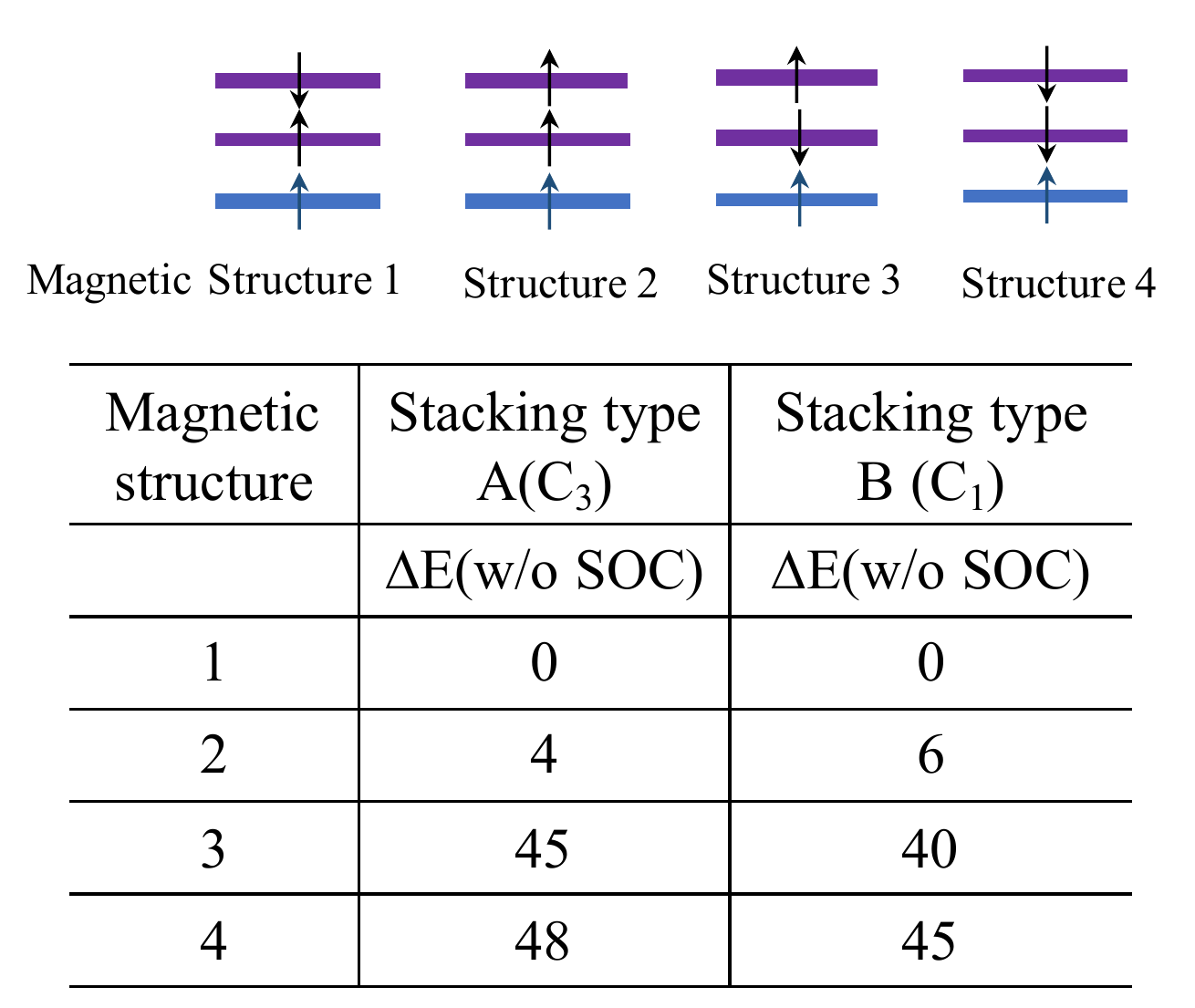}
\caption{\label{fig:2SLstacking}
The total energy $\Delta E$ for different magnetic structures of the interface for 2-SL \mbt on CrI$_3$. In each column the value of  $\Delta E$ is with respect to corresponding magnetic structure 1. One can find that the \mbt layer favors FM coupling with CrI$_3$ and AFM coupling to its neighboring \mbt. 
 }
\end{figure}

\begin{figure}[h]
\includegraphics[width=\linewidth]{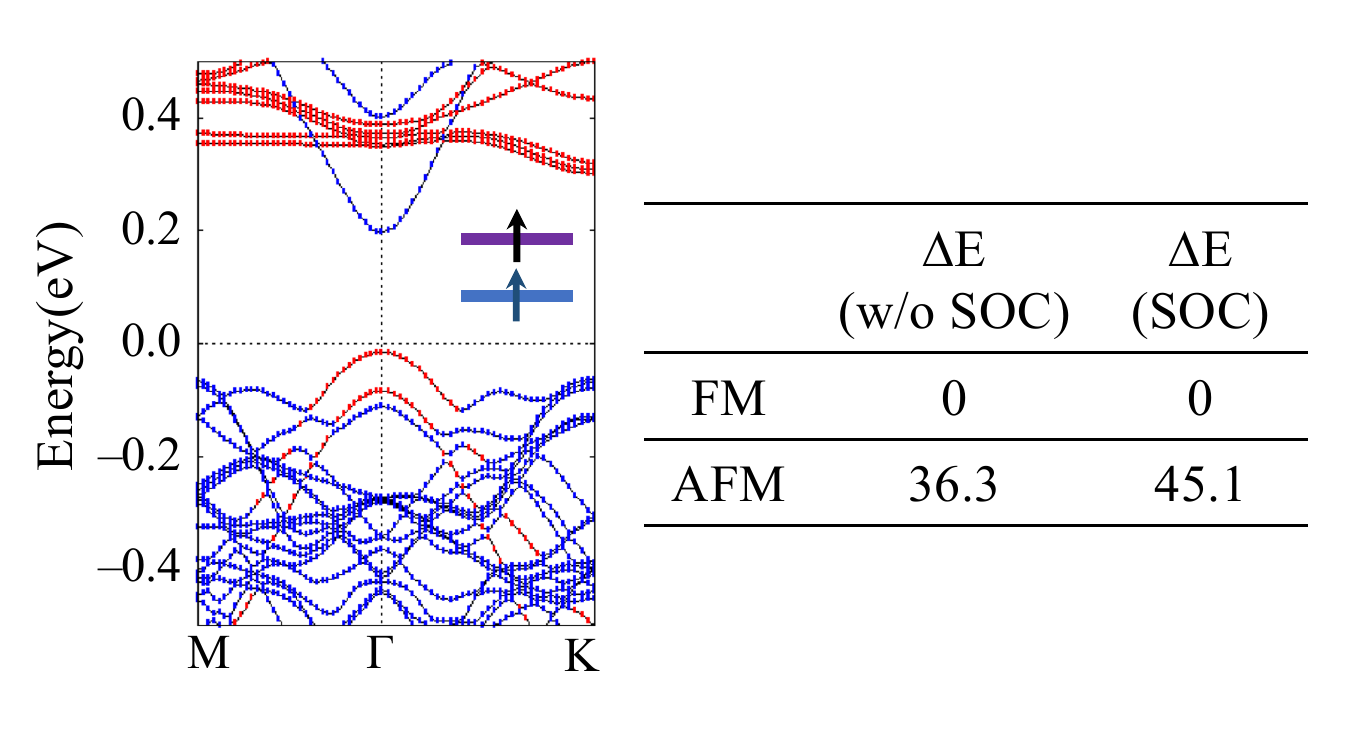}
\caption{\label{fig:3x3}
The total energy $\Delta E$ for different magnetic coupling of the interface for 1-SL $3\times3$ \mbt on $2\times2$ CrI$_3$. Here the magnetic coupling energy is still in the order of 40 meV and FM is favored in energy. For the FM coupling, the band structure exhibits an energy gap as a magnetic insulator. The red curves are Cr-$e_g$ bands and blue curves are \mbt states.}
\end{figure}

\begin{figure}[ht]
\includegraphics[width=\linewidth]{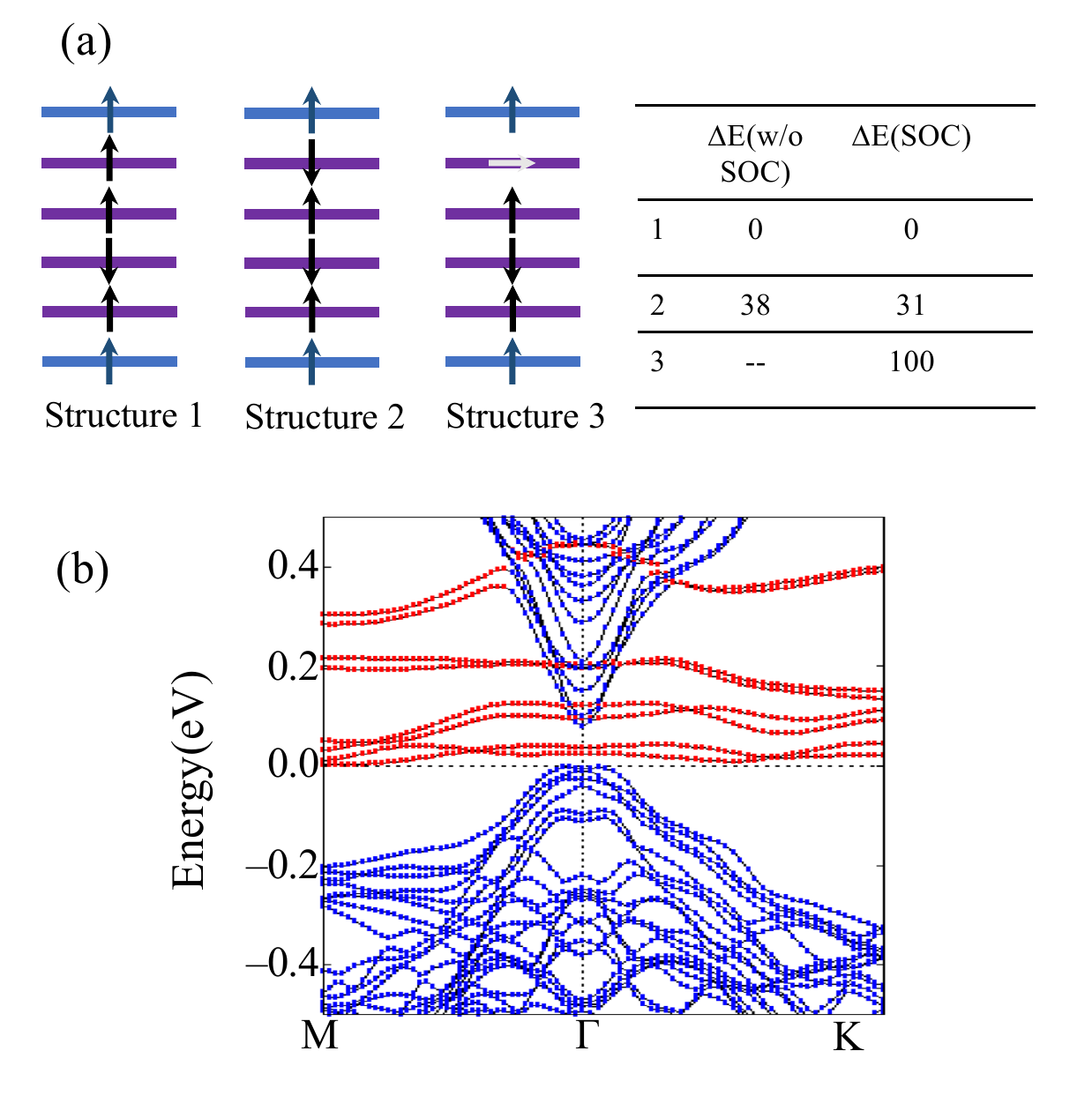}
\caption{\label{fig:4L-sandwich}
(a) Different magnetic structures for 4-SL-thick \mbt sandwiched between two CrI$_3$ layers and their total energies with respect to structure 1. 
We have tested different magnetic structures including the in-plane moment.
The top \mbt SL is switched to align parallel to the upper CrI$_3$ in spins, because of the exchange bias effect. 
(b) The band structure of structure 1. The red curves are Cr-$e_g$ bands and blue curves are \mbt states.
 }
\end{figure}

\begin{table}[h]
\caption{\label{tab:gap} The gap at $\Gamma$ from \mbt bands of the heterostructures.
When placing \mbt on top of CrI$_3$, the whole system remains to be insulating. Here, we show the band gap of the \mbt thin film as shown in Fig.~\ref{fig:band}.}
\begin{tabular}{c|c|c|c|c|c|c}
  \hline
Thickness (SL) &1  &2  &3  & 4 &5&6\\
 \hline
 Gap (meV)     &  299&59  &49  &42 &14&22  \\
 \hline
\end{tabular}
\end{table}

\begin{figure*}[h]
\includegraphics[width=\linewidth]{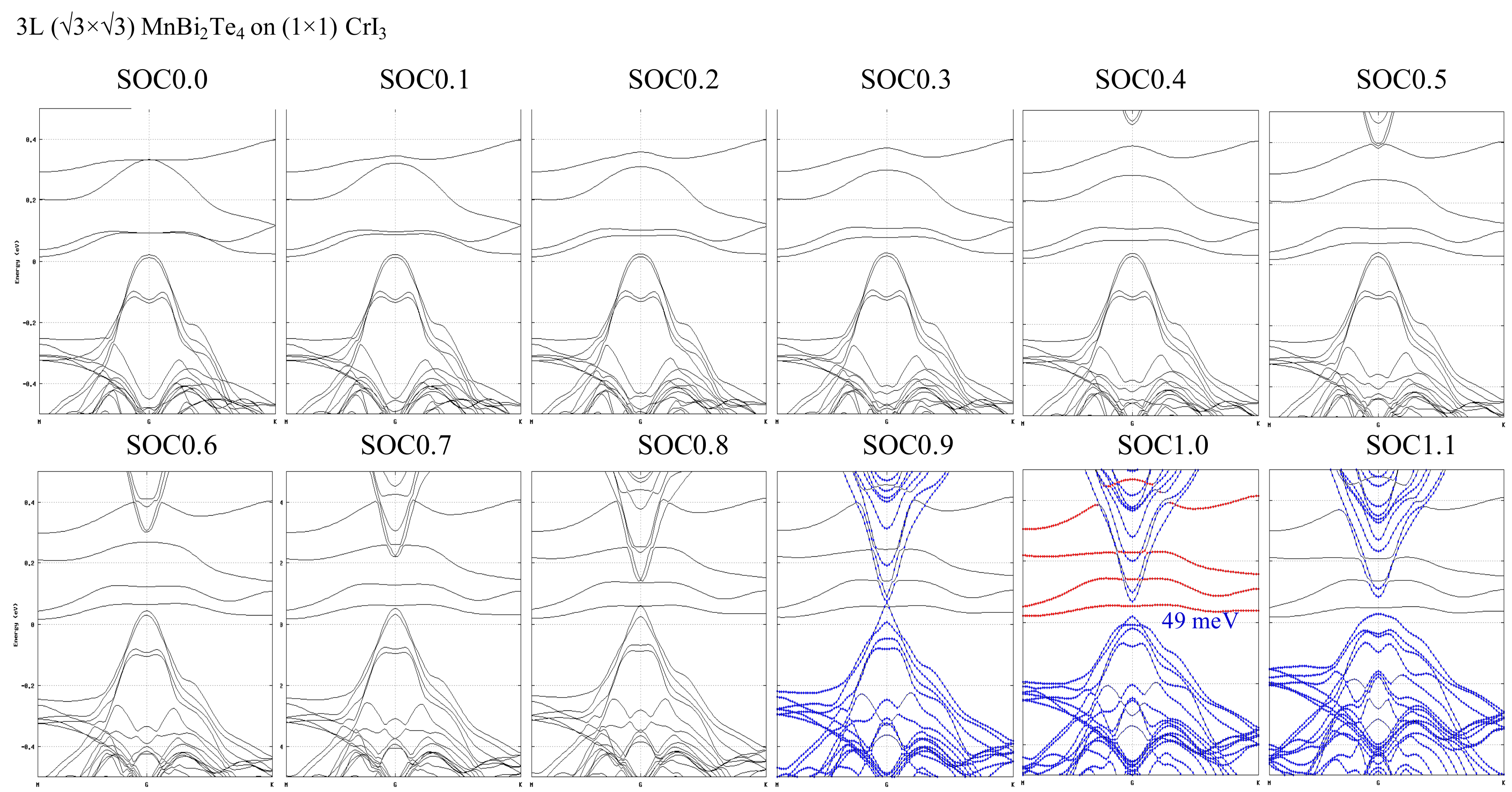}
\caption{\label{fig:3SL}
Band structure evolution for 3-SL-thick \mbt/CrI$_3$ heterostructure with varying SOC strength  from 0 to 110\%. 
There is a gap-closing near SOC =90\%, indicating a topological phase transition from a trivial magnetic insulator to a QAH insulator.
}
\end{figure*}

\begin{figure*}[h]
\includegraphics[width=\linewidth]{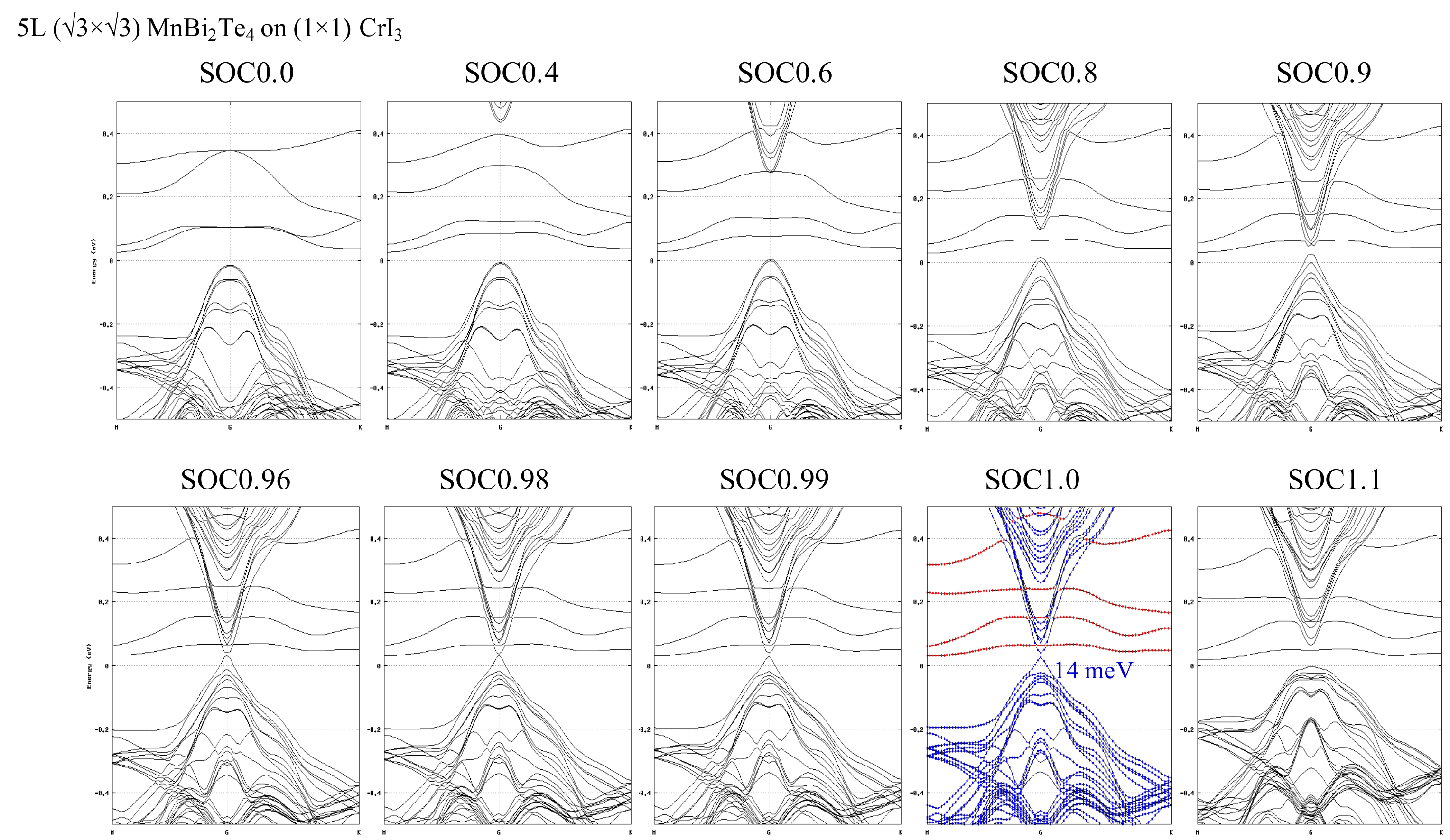}
\caption{\label{fig:5SL}
Band structure evolution for 5-SL-thick \mbt/CrI$_3$ heterostructure with varying SOC strength  from 0 to 110\%. 
There is a gap-closing near SOC =98\%, indicating a topological phase transition from a trivial magnetic insulator to a QAH insulator.
 }
\end{figure*}

\section{Structure models and exchange coupling}
In the interface model, we can overlap the CrI$_3$ layer and the \mbt layer with different sliding shift in the plane. We have tested different ways of stacking and found they give quantitatively very similar results in energy and band structure for the $\sqrt{3}\times \sqrt{3}$ MnBi$_2$Te$_4$/$1\times 1$ CrI$_3$ interface. At the same FM phase, the stacking way that preserves the three-fold rotation ($C_3$) exhibits slightly lower energy (by 20$\sim$30 meV for 1 to 3 \mbt SLs) than other low-symmetry stacking ways. Therefore, we used the $C_3$-symmetric model in the main text. 

We show the exchange coupling $\Delta_E = E_{FM} - E_{AFM}$  for 1-2 SL \mbt on a CrI$_3$ layer in  Figs.\ref{fig:1SLstacking} and \ref{fig:2SLstacking}.
Two different stacking ways give similar exchange coupling strength. One can also deduce that the \mbt - \mbt interlayer coupling ($\sim$ 4 meV) is much weaker than the \mbt - CrI$_3$ coupling ($\sim$ 40 meV) from Fig.\ref{fig:2SLstacking}.

For the much larger supercell, $3\times3$ MnBi$_2$Te$_4$/$2\times 2$ CrI$_3$, it still prefers the FM coupling at the interface with an exchange coupling energy $\sim$ 40 meV. The system remains as an insulator, as shown in Fig.~\ref{fig:3x3}.

The band gaps for different SLs of \mbt films (Fig.\ref{fig:band}) are shown in the table \ref{tab:gap}.

Figure \ref{fig:4L-sandwich} show the energies of different magnetic configurations of 4-SL of \mbt sandwiched between two CrI$_3$ layers. The structure 1, in which the top \mbt SL flips its spin direction due to the top CrI$_3$ layer, has the lowest energy. In the band structure, Cr-$e_g$ states appear as the conduction bands. There is still an energy gap for the QAH state with CN $= -1$, as shown in Fig.~\ref{fig:sandwich}.

Figures \ref{fig:3SL} and \ref{fig:5SL} show the band evolution as increasing SOC strength for 3 and 5 SL \mbt, respectively. The band gap closing is found when the SOC strength is increased to 90\% (98\%) of the full SOC strength for 3 (5) SL \mbt, reflecting the occurrence of TPT and the change of CN.

\end{document}